\documentclass[fleqn,usenatbib]{mnras}
\usepackage[T1]{fontenc}
\usepackage{ae,aecompl}

\usepackage{graphicx}	% Including figure files
\usepackage{amsmath}	% Advanced maths commands
\DeclareMathOperator{\sech}{sech}

\usepackage{amssymb}	% Extra maths symbols

\usepackage{txfonts} %old version of newtxtext, should be loaded after amsmath

\usepackage{epstopdf}
\usepackage{journals}

\usepackage[none]{hyphenat}
\graphicspath{{images/}}
\DeclareGraphicsExtensions{.pdf,.png,.jpg, .jpeg}
\usepackage{natbib}
\usepackage{color}
\usepackage{caption}
\usepackage{multirow}
\usepackage{threeparttable}

\title[Is the late buckling stage inevitable in the bar life?]{Is the late buckling stage inevitable in the bar life?}
\author[Anton A. Smirnov and Natalia Ya. Sotnikova]
{Anton A. Smirnov$^{1,2}$\thanks{E-mail:
zeleniikot@gmail.com (AAS)} and Natalia Ya. Sotnikova$^{1}$\\
$^{1}$St. Petersburg State University,
Universitetskij pr.~28, 198504 St. Petersburg, Stary Peterhof, Russia\\
$^{2}$Central (Pulkovo) Astronomical Observatory of RAS, Pulkovskoye Chaussee 65/1, 196140 St. Petersburg, Russia\\
}
\date{Accepted XXX. Received rYY; in original form ZZZ}

\pubyear{2018}

\begin{document}
%%%%%%%%%%%%%%%%%%%%%%%%%%%%%%%%%%%%%%%%%%%%%%%%%%%%%%%%%%%%%%%%%%%%%%%%%%%%%%%%%%%%%%%%%%%%%%%%%%%%%%%
\label{firstpage}
\pagerange{\pageref{firstpage}--\pageref{lastpage}}
\maketitle

%%%%%%%%%%%%%%%%%%%%%%%%%%%%%%%%%%%%%%%%%%%%%%%%%%%%%%%%%%%%%%%%%%%%%%%%%%%%%%%%%%%%%%%%%%%%%%%%%%%%%%%
\begin{abstract}
By means of self-consistent numerical simulations we investigated the dynamical impact of classical bulges on the growth of the secondary  buckling of a bar. Overall we considered 14 models with different disc and bulge parameters. We obtained that a bulge with a quite modest mass $B/D=0.1$ leads to completely symmetrical evolution of the bar almost independently of the initial stellar disc parameters and even can damp the first bending. At the same time, the bars in all our bulgeless models suffer from the short primary and prolonged secondary buckling. Given the smallness of the mass suppressing secondary buckling, we conclude that a classical bulge along with the gas central concentration may be the main culprits for the rarity of bars with ongoing buckling in the local Universe.    
\end{abstract}
%%%%%%%%%%%%%%%%%%%%%%%%%%%%%%%%%%%%%%%%%%%%%%%%%%%%%%%%%%%%%%%%%%%%%%%%%%%%%%%%%%%%%%%%%%%%%%%%%%%%%%%

%%%%%%%%%%%%%%%%%%%%%%%%%%%%%%%%%%%%%%%%%%%%%%%%%%%%%%%%%%%%%%%%%%%%%%%%%%%%%%%%%%%%%%%%%%%%%%%%%%%%%%%
\begin{keywords}
galaxies: bulges  -- galaxies: fundamental parameters -- galaxies: structure -- galaxies: kinematics and dynamics
\end{keywords}
%%%%%%%%%%%%%%%%%%%%%%%%%%%%%%%%%%%%%%%%%%%%%%%%%%%%%%%%%%%%%%%%%%%%%%%%%%%%%%%%%%%%%%%%%%%%%%%%%%%%%%%

%%%%%%%%%%%%%%%%%%%%%%%%%%%%%%%%%%%%%%%%%%%%%%%%%%%%%%%%%%%%%%%%%%%%%%%%%%%%%%%%%%%%%%%%%%%%%%%%%%%%%%%
\section{Introduction}
%%%%%%%%%%%%%%%%%%%%%%%%%%%%%%%%%%%%%%%%%%%%%%%%%%%%%%%%%%%%%%%%%%%%%%%%%%%%%%%%%%%%%%%%%%%%%%%%%%%%%%%
Galactic discs are very fragile systems. They are highly responsive to the external gravitational impact and even to small internal perturbations which excite global in-plane density waves (a bar, a spiral pattern) and corrupts the disc in the vertical direction via bending modes creating extended S-shape or U-shape warps. Galactic bars are also the subject to the short-term and violent bending/buckling instability leading to the formation of so-called boxy/peanut shaped (B/PS) structures, which are well recognised if the galaxy is seen edge-on \citep{Lutticke_etal2000}. 
\par
Unfortunately, the main evidence of the bar sensitivity to the buckling comes from simulations \citep{Combes_etal1990,Raha_etal1991,Pfenniger_Friedli1991,Berentzen_etal1998,Oneil_Dubinski2003,MartinezValpuesta_etal2004,Debattista_etal2006,MartinezValpuesta_etal2006,Saha_etal2013}, the stability analysis of bending modes (e.g., \citealp{Sotnikova_Rodionov2003}) and studies of orbital dynamics in bar toy models (e.g., \citealp{Combes_etal1990,Patsis_etal2002,Patsis_Katsanikas2014}). The real observation statistics of buckling bars is very poor. Currently, only three galaxies at an intermediate inclination (NGC~3227, NGC~4569 and ESO~506--G004) show indirect signs of \textcolor{black}{ongoing} bar buckling \citep{Erwin_Debattista2016,Li_etal2017} and it is unclear how widespread they are in the Universe. 
\par
Early numerical studies of the bar vertical evolution showed that the bar bends (buckles) \textcolor{black}{out of the plane} fairly quickly after it has been formed \textcolor{black}{and becomes asymmetric with respect to the equatorial plane} 
\citep{Friedli_Pfenniger1990,Raha_etal1991,Pfenniger_Friedli1991,Sotnikova_Rodionov2003,MartinezValpuesta_etal2004}. The typical time scale between bar formation and buckling times is about 1-2 Gyr and the buckling phase itself is very rapid, i.e. about a few hundred Myr (see \citet{Athanassoula2016} for a review and especially figure 3 therein). Since the entire time period is relatively short, this somehow explained the rarity of the \textcolor{black}{buckled} bars. Estimates of the frequency of \textcolor{black}{buckled} bars \citep{Erwin_Debattista2016} were based on the existence of only primary (early) buckling but the problem is actually more serious because of the so-called recurrent buckling\footnote{The term ``recurrent buckling'' was introduced by \citet{MartinezValpuesta_etal2006} but it means just a secondary episode.}. \citet{MartinezValpuesta_etal2006} showed that bars could buckle (lose the vertical symmetry) the second time. \textcolor{black}{In simulations by} \citet{MartinezValpuesta_etal2006} secondary buckling occurred after the first one, about $6-7$ Gyr from the start of simulations, and the bar remained vertically asymmetric for a prolonged (around $3$ Gyr) time interval. Moreover, by this time the bar had grown in size and the degree of its vertical asymmetry was significant \textcolor{black}{and strongest in the middle of the bar region}. \textcolor{black}{In hindsight, such events can also be seen in other simulations} \citep{Oneil_Dubinski2003,Athanassoula2005b}. \textcolor{black}{One model (RCG051A) by \citet{Saha_etal2013} with a very light bulge also demonstrates at least two buckling episodes. Primary buckling was early and short and second one was later and prolonged}.
\par
\textcolor{black}{The existence of the later and prolonged buckling stage of a bar creates a problem for interpreting observational data.}  \citet{MartinezValpuesta_etal2006} noted that, although in principle it is possible \textcolor{black}{to detect observationally the buckled discs} 
%the bar remains vertically asymmetric to observe recurrent buckling, 
there are two factors that can suppress \textcolor{black}{the buckling at all, both earlier and later}.
%it. 
One of them is well-studied in the literature and is associated with the presence of the gas component. Numerical simulations \citep{Berentzen_etal1998,Debattista_etal2006,Berentzen_etal2007} have shown that if gas physics allows gas to sink into the centre of the disc, increasing the central concentration, the buckling instability can be completely suppressed, and the vertical bar evolution proceeds symmetrically. The presence of gas explains why the late-type galaxies do not possess ``buckled'' bars with boxy or barlens isophotes \textcolor{black}{in the bar plane}  \citep{Erwin_Debattista2017,Li_etal2017}, which are candidates for the ongoing \textcolor{black}{vertically} buckling bars. As for the  early-type galaxies, in which there is not much gas, buckled bars could exist, however, again, this is not confirmed by observations, and the problem remains. \textcolor{black}{The second factor mentioned by} \citet{MartinezValpuesta_etal2006} is ``unfavorable initial conditions'' for the secondary buckling (see Conclusions in \citealp{MartinezValpuesta_etal2006}). \textcolor{black}{Unfortunately, the authors did not consider them in detail}. 
\par
\citet{Smirnov_Sotnikova2018} showed that, in contrast, ``favorable conditions'' are quite common, \textcolor{black}{at least in models with no gas}. For example, the cool stellar isothermal disc with the vertical/radial scale length ratio $Z_\mathrm{d}/R_\mathrm{d}=0.1$ and with the NWF \citep{NFW} dark halo profile is subjected to prolonged secondary buckling \textcolor{black}{lasting over 4 Gyr}! Moreover, our simulations of a large number of models with different initial conditions showed that secondary prolonged buckling is almost a natural turn of events for simulated galaxies, with the exception of some specific models. There were only two groups of models where the \textcolor{black}{buckling} was completely damped: one group with \textcolor{black}{heavy} classical bulges and one model with a heavy dark halo\footnote{\citet{Saha_etal2013} give an example of the model with a very heavy halo that does not experiences buckling at all but demonstrates a bar and a weakly expressed B/PS structure; model RHG057.} ($M_\mathrm{h}(R<4R_\mathrm{d})/M_\mathrm{d} \ga 3$). In other words, spherical subsystems consisting of stars or dark matter contribute to the central concentration and tend to weaken the bending instability just as the gas component does. \citet{Sotnikova_Rodionov2005} made such a conclusion analysing the modified dispersion equation for bending modes \citep{Sellwood1996} and confirmed it by numerical experiments. However, a few models considered in \citet{Sotnikova_Rodionov2005} or \citet{Smirnov_Sotnikova2018} do not allow to estimate the real impact of spherical subsystems on the \textcolor{black}{buckling}. Is it so strong that any bending modes turn out to be completely damped or will the \textcolor{black}{loss of the vertical symmetry} still occur but at \textcolor{black}{the very late} stages of evolution? It is important to answer this question in order to understand where to look for the galaxies with buckling (asymmetric in the vertical direction) bars.
\par 
In this article, we study the impact of a classical bulge on the bar buckling. \textcolor{black}{First of all, we are interested in the conditions under which the secondary late buckling is suppressed.}  We perform a set of numerical simulations with high spatial resolution (Section~2). We employ a series of numerical bulgeless galaxy models from our previous work \citep{Smirnov_Sotnikova2018}. These models demonstrate clear \textcolor{black}{secondary} buckling event. For each of them, we construct a counterpart with the same initial parameters but with the addition of a small bulge component. Both types of models were simulated for about 8 Gyr (Section~3). We obtained that even the addition of a very low-mass component ($M_\mathrm{b} = 0.1 M_\mathrm{d}$) damped the secondary buckling throughout all the time of simulations. For one family of models, we found the boundary bulge mass $M_\mathrm{b} \approx 0.05 M_\mathrm{d}$, for which the secondary buckling would still manifests itself (Section~4). Even such a small classical bulge component, which is hardly detected by decomposition, can be initially nested in the disc and can lead to the symmetric vertical evolution of a bar. Based on these results, we conclude that classical bulges may be the reason why almost all bars in early-type galaxies do not exhibit a buckling phase (Section~5).

%%%%%%%%%%%%%%%%%%%%%%%%%%%%%%%%%%%%%%%%%%%%%%%%%%%%%%%%%%%%%%%%%%%%%%%%%%%%%%%%%%%%%%%%%%%%%%%%%%%%%%%
\section{Numerical model}
%%%%%%%%%%%%%%%%%%%%%%%%%%%%%%%%%%%%%%%%%%%%%%%%%%%%%%%%%%%%%%%%%%%%%%%%%%%%%%%%%%%%%%%%%%%%%%%%%%%%%%%

%%%%%%%%%%%%%%%%%%%%%%%%%%%%%%%%%%%%%%%%%%%%%%%%%%%%%%%%%%%%%%%%%%%%%%%%%%%%%%%%%%%%%%%%%%%%%%%%%%%%%%%
\begin{table}
\centering
\caption{Parameters of models}
\begin{tabular}{ c | c | c | c | c | c | c |}
\hline
$M_\mathrm{h} (r < 4R_\mathrm{d})$& $z_\mathrm{d} / R_\mathrm{d}$ & $Q$ & $M_\mathrm{b}$ & $r_\mathrm{b}$ &$N_\mathrm{b}$, \textit{kk}\\
\hline 
\hline
 1.0 & 0.05 & 1.2&0& -- & -- \\
 \hline
 1.0 & 0.05 & 1.2&0.2& 0.2 & 0.8  \\
\hline
 1.5 & 0.05  &  1.2 &0& ---&  --- \\
 \hline
 1.5 & 0.05  &  1.2 & 0.2 & 0.2  & 0.8\\
\hline
 1.5 & 0.05  &  1.6 &0& ---&  --- \\
 \hline
 1.5 & 0.05  &  1.6 & 0.2 & 0.2  & 0.8\\
 \hline
 1.5 & 0.1  &  1.2 &0& ---&  --- \\
 \hline
 1.5 & 0.1  &  1.2 & 0.2 & 0.2  & 0.8\\
\hline
 1.5 & 0.2  &  1.2 & 0 & ---&  --- \\
 \hline
 1.5 & 0.2  &  1.2 & 0.05 & 0.4  & 0.2\\
 \hline
 1.5 & 0.2  &  1.2 & 0.05 & 0.2  & 0.2\\
 \hline
 1.5 & 0.2  &  1.2 & 0.1 & 0.2  & 0.4\\
 \hline
 1.5 & 0.2  &  1.2 & 0.2 & 0.2  & 0.8\\
 \hline
 1.5 & 0.2  &  1.2 & 0.3 & 0.2  & 1.2\\
\hline

%%%%%%%%%%%%%%%%%%%%%%%%%%%%%%%%%%%%%%%%%%%%%%%%%%%%%%%%%%%%%%%%%%%%%%%%%%%%%%%%%%%%%%%%%%%%%%%%%%%%%%%
\multicolumn{7}{p{0.4\textwidth}}
{\footnotesize{\textit{Notes}: each column represents parameters of the models, one model on one line. $M_\mathrm{h}(R < 4R_\mathrm{d})$ is the mass of the halo in units of the disc mass $M_\mathrm{d}$ within a sphere with radius $R=4R_\mathrm{d}$, where $R_\mathrm{d}$ is the scale length of the disc, $z_\mathrm{d}/R_\mathrm{d}$ is the initial ratio of the disc scale height to the disc scale length. $M_\mathrm{b}$ and $r_\mathrm{b}$ are the total mass and the scale length of the bulge, respectively. $N_\mathrm{b}$ is the number of particles in the bulge.}}
\end{tabular}
\label{tab:models_pars}
\end{table} 
%%%%%%%%%%%%%%%%%%%%%%%%%%%%%%%%%%%%%%%%%%%%%%%%%%%%%%%%%%%%%%%%%%%%%%%%%%%%%%%%%%%%%%%%%%%%%%%%%%%%%%%
By means of self-consistent numerical simulations \citet{Smirnov_Sotnikova2018} showed that different values of initial parameters of the galaxy model (disc thickness, Toomre parameter, dark halo mass, presence of a bulge) lead to different scenarios of the bar evolution in the vertical direction and different kinds of the secondary buckling. To study the bulge impact on the buckling we construct counterparts with a bulge to each bulgeless model with \textcolor{black}{secondary} buckling from \citet{Smirnov_Sotnikova2018}.  Overall, we consider fourteen models. For the sake of consistency, all new models were constructed in the same way as it was done in \citet{Smirnov_Sotnikova2018}.  \par 
Each model consists of \textcolor{black}{a radially exponential and vertically isothermal disc} with a radial scale $R_\mathrm{d}$, \textcolor{black}{vertical scale $Z_\mathrm{d}$} and the total mass $M_\mathrm{d}$:
\begin{equation}
\rho(r,z) = \frac{M_\mathrm{d}}{4\pi R_\mathrm{d}^2 z_\mathrm{d}} \cdot \exp(-R/R_\mathrm{d}) \cdot \sech^2(z/z_\mathrm{d}) \,,
\label{eq:sigma_disc} 
\end{equation}
a dark halo of the NFW type \citep{NFW}:
%%%%%%%%%%%%%%%%%%%%%%%%%%%%%%%%%%%%%%%%%%%%%%%%%%%%%%%%%%%%%%%%%%%%%%%%%%%%%%%%%%%%%%%%%%%%%%%%%%%%%%%
\begin{equation}
\rho = 
\frac{C_\mathrm{h}\,T(r/r_\mathrm{t})}
{(r/r_\mathrm{s})^{\gamma_0}
\left((r/r_\mathrm{s})^{\eta}+1\right)^
{(\gamma_{\infty}-\gamma_0)/\eta}} \,,
\label{eq:NFW}
\end{equation}
%%%%%%%%%%%%%%%%%%%%%%%%%%%%%%%%%%%%%%%%%%%%%%%%%%%%%%%%%%%%%%%%%%%%%%%%%%%%%%%%%%%%%%%%%%%%%%%%%%%%%%%
where $r_\mathrm{s}$ is the halo scale radius, $r_\mathrm{t}$ is the halo truncation radius, $\eta$ is the halo transition exponent, $\gamma_0$ is the halo inner logarithmic density slope, $\gamma_{\infty}$ is the halo outer logarithmic density slope, $C_\mathrm{h}$ is the parameter defining the full mass of the halo $M_\mathrm{h}$ and $T(x)$ is the truncation function: 
%%%%%%%%%%%%%%%%%%%%%%%%%%%%%%%%%%%%%%%%%%%%%%%%%%%%%%%%%%%%%%%%%%%%%%%%%%%%%%%%%%%%%%%%%%%%%%%%%%%%%%%
\begin{equation}
T(x) = \frac{2}{\sech{x} + 1/\sech{x}} \,.
\end{equation}
%%%%%%%%%%%%%%%%%%%%%%%%%%%%%%%%%%%%%%%%%%%%%%%%%%%%%%%%%%%%%%%%%%%%%%%%%%%%%%%%%%%%%%%%%%%%%%%%%%%%%%%
We use $\eta=4/9$, $\gamma_0=7/9$, $\gamma_{\infty}=31/9.$
If a bulge component is present it is defined by a Hernquist profile \citep{Hernquist1990}:
%%%%%%%%%%%%%%%%%%%%%%%%%%%%%%%%%%%%%%%%%%%%%%%%%%%%%%%%%%%%%%%%%%%%%%%%%%%%%%%%%%%%%%%%%%%%%%%%%%%%%%%
\begin{equation}
\rho_\mathrm{b} = \frac{M_\mathrm{b}\, r_\mathrm{b}}{2\pi\,r\,(r_\mathrm{b} + r)^3} \,, 
\end{equation} 
%%%%%%%%%%%%%%%%%%%%%%%%%%%%%%%%%%%%%%%%%%%%%%%%%%%%%%%%%%%%%%%%%%%%%%%%%%%%%%%%%%%%%%%%%%%%%%%%%%%%%%%
where $r_\mathrm{b}$ is the scale parameter and $M_\mathrm{b}$ is the total bulge mass. We use a large number of particles to represent each component. The disc and the halo consist of $4kk$ and $4.5kk$ particles, respectively. Number of particles in the bulge is determined in such manner that the mass of one particle from the disc must be equal to the mass of one particle from the bulge. Tab.~\ref{tab:models_pars} summarises important  details of model parameters. For the bulge parameters, we stick them to quite typical observed values of mass and scale length, $M_\mathrm{b}=0.2$ and $r_\mathrm{b}=0.2$ (e.g., \citealp{Mosenkov_etal2010}). We consider discs of various thickness, from $Z_\mathrm{d}/R_\mathrm{d}=0.05$ to $0.2$. We use two values of the Toomre parameter, $Q=1.2$ and $Q=1.6$. These values determine the constant $\sigma_0 $ in the initial velocity dispersion profile $\sigma_R$ of the disc, which is exponential with a typical scale twice of the density scale
%%%%%%%%%%%%%%%%%%%%%%%%%%%%%%%%%%%%%%%%%%%%%%%%%%%%%%%%%%%%%%%%%%%%%%%%%%%%%%%%%%%%%%%%%%%%%%%%%%%%%%%
\begin{equation}
\sigma_R = \sigma_0 \cdot \exp(-R/2 R_\mathrm{d}).
\end{equation}
%%%%%%%%%%%%%%%%%%%%%%%%%%%%%%%%%%%%%%%%%%%%%%%%%%%%%%%%%%%%%%%%%%%%%%%%%%%%%%%%%%%%%%%%%%%%%%%%%%%%%%%
For one model with a thick disc $Z_\mathrm{d}=0.2$\footnote{We call this disc ``thick'' but the statistics of the relative disc thickness give the ratio $Z_\mathrm{d}/R_\mathrm{d}=0.2$ as a typical value for edge-on galaxies \citealp{Mosenkov_etal2015}.}, we use an extended list of the bulge initial parameters and look for the boundary mass of the bulge, which can no longer prevent secondary buckling.
\par 
The initial equilibrium state is prepared via a script for constructing the equilibrium multicomponent model of a galaxy {\tt{mkgalaxy}} \citep{McMillan_Dehnen2007} from the toolbox for N-body simulation {\tt{NEMO}} \citep{Teuben_1995}. 
The equations of motions were solved by the built-in fast numerical integrator {\tt{gyrfalcON}} \citep{Dehnen2002}. The simulation time for all models was about 8 Gyr. 
\textcolor{black}{We scale the disc parameters as $M_\mathrm{d} = M_\mathrm{u} = 1$, $R_\mathrm{d} = R_\mathrm{u} = 1$ with $G = 1$ 
and keep in mind the typical values for these parameters $M_\mathrm{d} = 10^{10} M_{\sun}$, $R_\mathrm{d} = 3.5$~kpc.}
%with an integration time-step of about 13 Myr. 
For each individual run the sum of potential and kinetic energy was conserved with an accuracy of less than $0.1 \%$.\\
%(The construction of self-consistent galaxy multicomponent models actually is  quite a simple task nowadays if you have appropriate hardware and have a programmer to install appropriate software. There exist several numerical packages which can give you a desired model after pressing just one magic button.)
%%%%%%%%%%%%%%%%%%%%%%%%%%%%%%%%%%%%%%%%%%%%%%%%%%%%%%%%%%%%%%%%%%%%%%%%%%%%%%%%%%%%%%%%%%%%%%%%%%%%
\section{Results}
%%%%%%%%%%%%%%%%%%%%%%%%%%%%%%%%%%%%%%%%%%%%%%%%%%%%%%%%%%%%%%%%%%%%%%%%%%%%%%%%%%%%%%%%%%%%%%%%%%%%%%%
\begin{figure*}
\begin{minipage}[t]{0.49\textwidth}%
\includegraphics[scale=0.3]{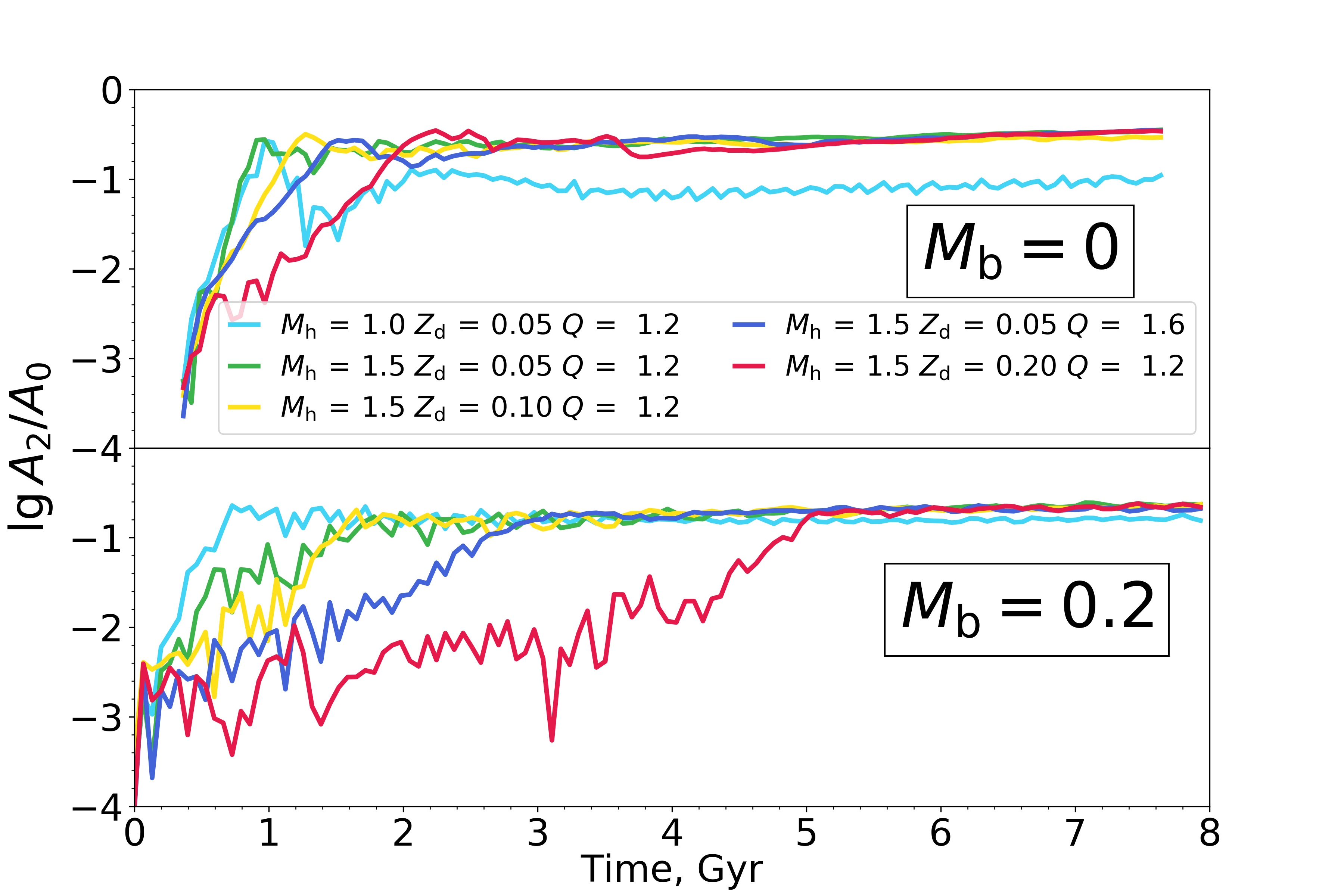}
%\label{fig:bar_speed}
\end{minipage}%
\begin{minipage}[t]{0.49\textwidth}%
\includegraphics[scale=0.3]{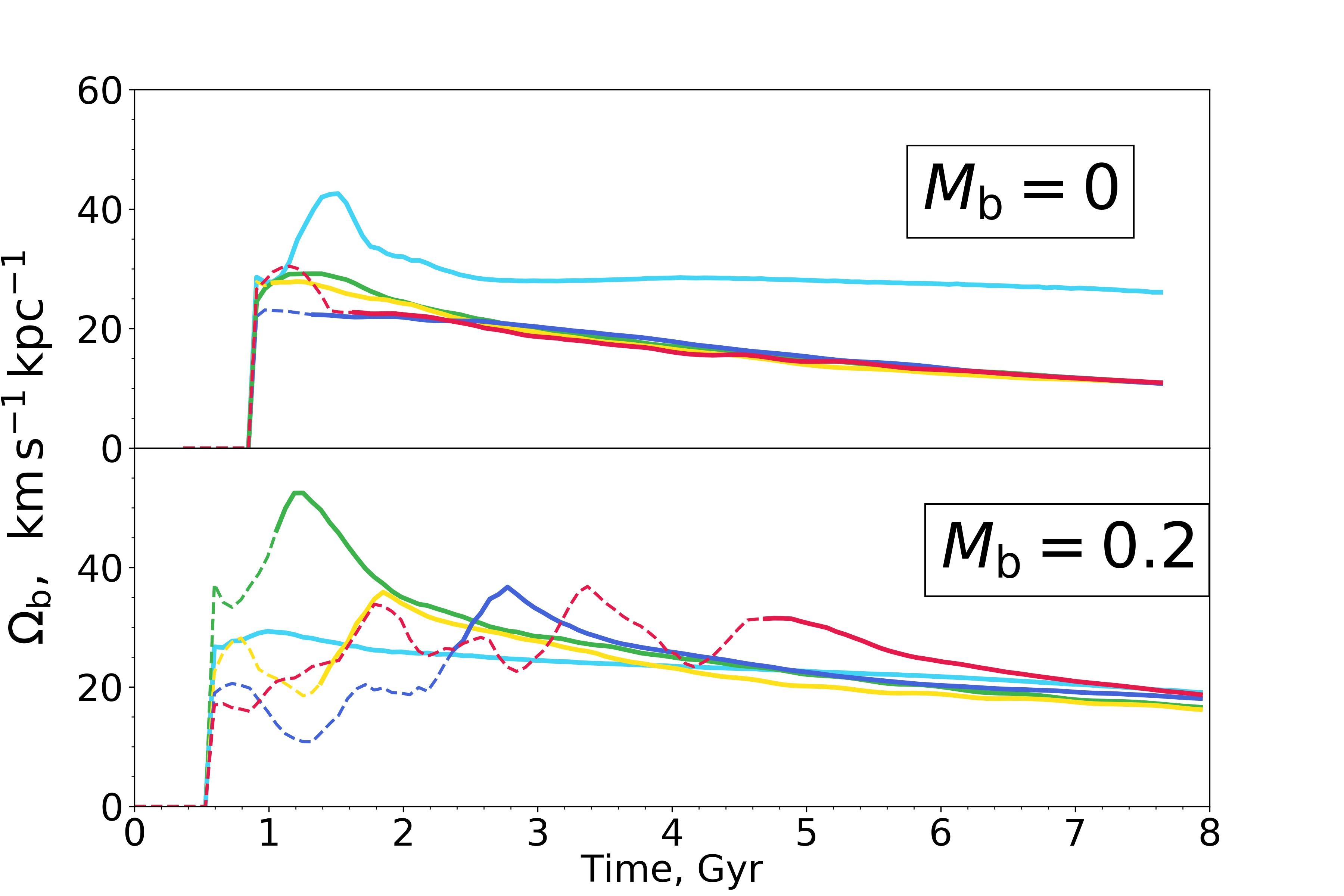}
%\label{fig:axis}
\end{minipage}%
\caption{The decimal logarithm of the normalised amplitude ({\it left}) and the pattern speed ({\it right}) of the bar in models without and with bulges (top and lower panels respectively) and different initial parameters of the stellar disc. Dotted lines on the right graph correspond to the period of time when the bar is not yet formed, or $A_\mathrm{2}/A_\mathrm{0}<0.05$.}
\label{fig:ampl}
\end{figure*}
%%%%%%%%%%%%%%%%%%%%%%%%%%%%%%%%%%%%%%%%%%%%%%%%%%%%%%%%%%%%%%%%%%%%%%%%%%%%%%%%%%%%%%%%%%%%%%%%%%%%%%%

%%%%%%%%%%%%%%%%%%%%%%%%%%%%%%%%%%%%%%%%%%%%%%%%%%%%%%%%%%%%%%%%%%%%%%%%%%%%%%%%%%%%%%%%%%%%%%%%%%%%
\begin{figure}
\begin{minipage}[t]{0.9\textwidth}%
\includegraphics[scale=0.3]{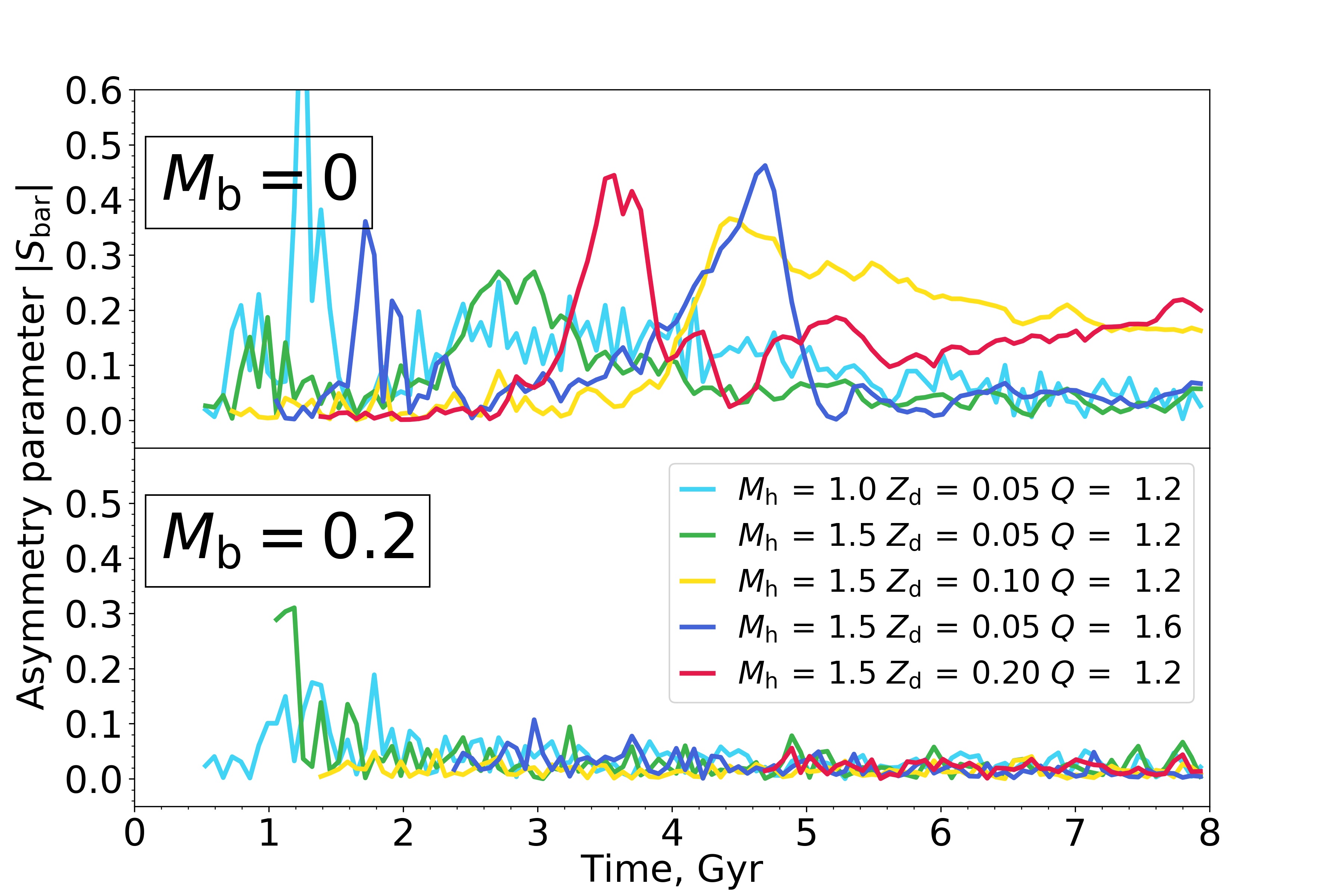}
%\label{fig:axis}
\end{minipage}%
\caption{Absolute value of the asymmetry parameter $|S_\mathrm{bar}|$ (see text for details) for models without and with bulges (upper and lower panels respectively) and different initial parameters of the stellar disc. \textit{Note}: All lines are drawn from the moment of bar formation or $A_\mathrm{2}/A_\mathrm{0} \gtrsim 0.05$ in the corresponding model (see text for details).}
\label{fig:asym_big}
\end{figure}
%%%%%%%%%%%%%%%%%%%%%%%%%%%%%%%%%%%%%%%%%%%%%%%%%%%%%%%%%%%%%%%%%%%%%%%%%%%%%%%%%%%%%%%%%%%%%%%%%%%%%%%

%%%%%%%%%%%%%%%%%%%%%%%%%%%%%%%%%%%%%%%%%%%%%%%%%%%%%%%%%%%%%%%%%%%%%%%%%%%%%%%%%%%%%%%%%%%%%%%%%%%%
\begin{figure*}
\begin{minipage}[t]{1.0\textwidth}%
\includegraphics[width=\textwidth]{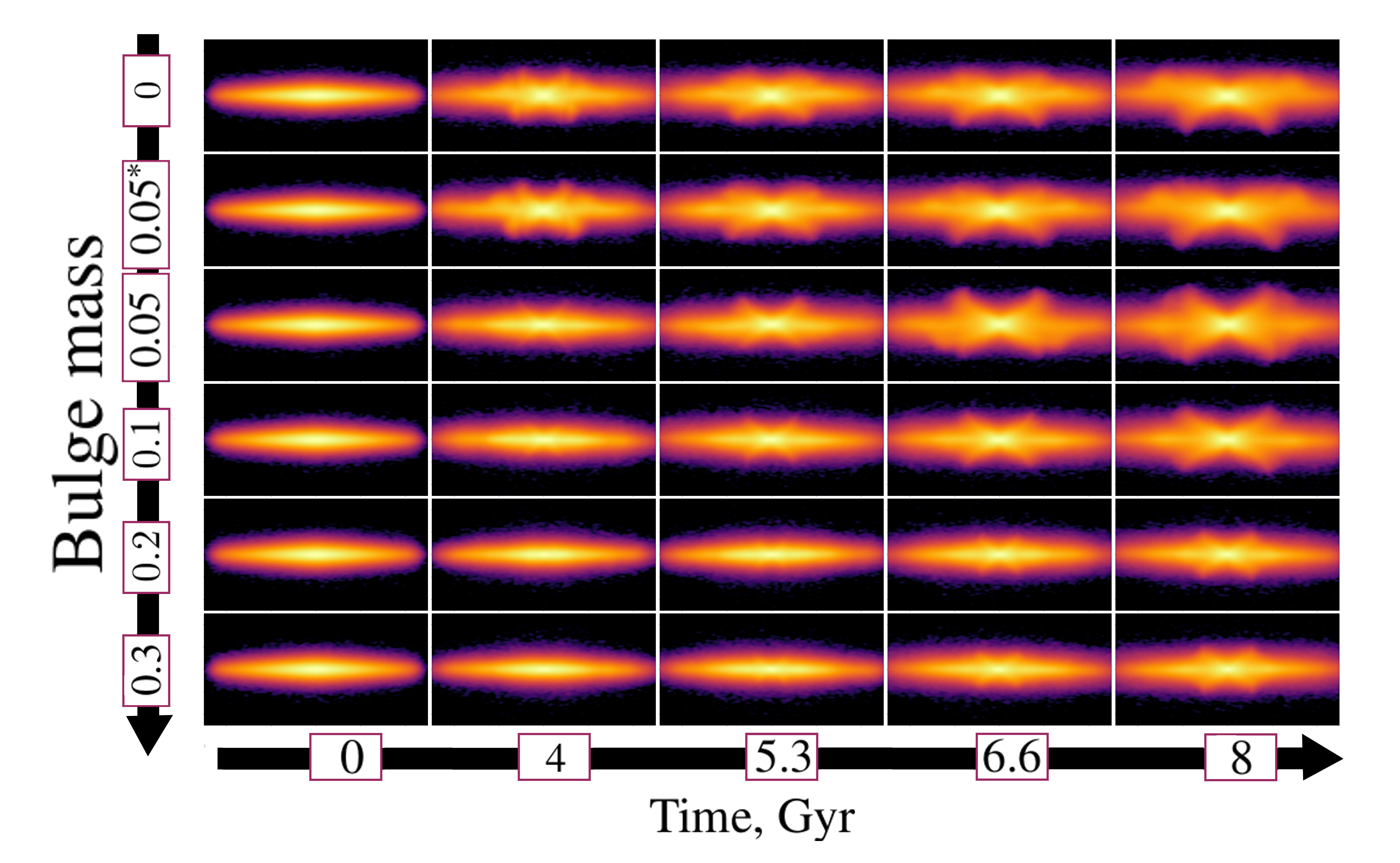}
\label{fig:axis}
\end{minipage}%
\caption{An edge-on view of models with different bulge contribution for different times; the asterisk symbol  denotes a model with a bulge scale length $r_\mathrm{b}=0.4$. The central concentration increases from top to bottom, time increase from left to right. With an increase of central concentration the bar loses its asymmetrical appearance (see upper right corner). \textcolor{black}{The size of each rectangle is 5x2.5 in our unit of the length $R_\mathrm{u}$.}}
\label{fig:model_snapshots}
\end{figure*}
%%%%%%%%%%%%%%%%%%%%%%%%%%%%%%%%%%%%%%%%%%%%%%%%%%%%%%%%%%%%%%%%%%%%%%%%%%%%%%%%%%%%%%%%%%%%%%%%%%%%%%%
All models follow the same evolutionary pattern, which is often observed in numerical simulations (e.g., \citealp{Athanassoula2005a}). The initially symmetric state is disturbed by small density waves,  which gradually grow, overlap at some point in time and then the forming bar becomes visible. 
Notably, the models with bulges tend to have a longer bar formation time which is consistent with previous studies of the bulge impact on the in-plane bar morphology \citep{Kataria_Das2018,Saha_Elmegreen2018}. \par
The amplitudes and  pattern speed of the bar for all models are shown in Fig.~\ref{fig:ampl}. By 8 Gyr, the pattern speed is grouped at close values for bulgeless galaxies and for galaxies with bulge, \textcolor{black}{regardless of the initial disc thickness and the parameter $Q$s}.  
For models with $M_\mathrm{b}=0.2$, the bar is formed about 1.5~Gyr later. The thick model shows an even greater delay in the bar formation.
\par
\textcolor{black}{Several definitions have been proposed and used to determine the strength of the buckling event and the time at which it occurs \citep{Merritt_Sellwood1994,MartinezValpuesta_etal2006,Debattista_etal2006,Berentzen_etal2007,MV_Athanassoula2008,Saha_etal2013}. Among other methods, the buckling instability can be indicated by a drop in $A_2$ or $\sigma_z/\sigma_R$ with time. One can also measure the strength of the buckling measuring the mean ($< z >$) \citep{MV_Athanassoula2008}. Most often, the strength of the buckling amplitude by computing $m=1$ Fourier component $A_{1z}$ in the $xz$-plane of the disc where the major axis of the bar is oriented along the $x$ axis \citep{MV_Athanassoula2008}.
We tried this quantity to account for the vertical asymmetry, but it turned out to be quite noisy if all disc particles are processed. The use of these parameters requires the selection of a specific area confining a bar to reduce noise.}
To characterise the buckling strength we employ the asymmetry parameter which is calculated in the following way \citep{Smirnov_Sotnikova2018}: $S_\mathrm{bar} = (A_2(z>0)-A_2(z<0))/A_2$, where $A_2(z>0)$ is the amplitude of the bar above the disc plane ($z>0$), whereas $A_2(z<0)$ is the amplitude below the disc plane ($z<0$). 
The advantage of using our parameter is that even if we process all disc particles the noise level is rather low (unless the bar amplitude is near zero) and therefore it is more appropriate for different galaxy models where different bars arise. \textcolor{black}{The other side of the coin is that the parameter is not appropriate for estimating the amplitude of bending modes which do not affect the bar, i. e. small-scale bending waves arising at initial stages of model evolution. In the scope of the present work we are mainly interested in late and prolonged bar buckling stages and therefore the use of our parameter is more than justified.}  \par  %Not to mention the simplicity of calculation.
%\textcolor{black}{To simplify graphs with a large number of curves we use the absolute value of the asymmetry parameter $|S_\mathrm{bar}|$ .} 
\textcolor{black}{Bars in our models can be bent either to the north or the south poles equally, without a preferred direction. Therefore without the loss of generality and to simplify further discussion we consider only the absolute values of asymmetry parameter $|S_\mathrm{bar}|$.}  Fig.~\ref{fig:asym_big} shows \textcolor{black}{the absolute value of the asymmetry parameter} as a function of time for models without bulge and for the same models with a bulge with $M_\mathrm{b}=0.2$ and $r_\mathrm{b}=0.2$. \textcolor{black}{For the sake of simplicity we plot all dependencies from the moment of bar formation}. Prior to this, the asymmetry shows large fluctuations with no significant physical meaning as bar amplitude $A_2$ is close to zero (see Fig.~\ref{fig:ampl}) = small numbers in the denominator in $S_\mathrm{bar}$ equation.
%The bar is not yet formed, the amplitude $A_2$ is close to zero (see Fig.~\ref{fig:ampl}) = small numbers in the denominator in $S_\mathrm{bar}$ equation which cause large amplitude variations. This phase ends as soon as the bar has formed.
We distinguish a bar in our models when $A_2$ is greater than approximately $5 \cdot 10^{-2}$. As a rule, a bar is formed in 1-2~Gyr but, in some cases ($Q=2.0$), it happens in about 3-4~Gyr. The further evolution is quite dependent on initial model parameters. Immediate striking result that follows from the figure is that all the models with a bulge have a vertically symmetric bar (\textcolor{black}{$|S_\mathrm{bar}| \approx 0$}) while their bulgeless counterparts suffer from the \textcolor{black}{secondary} buckling (\textcolor{black}{$|S_\mathrm{bar}|$ is noticeably greater than zero.}). The effect of the bulge manifests itself most clearly in the model with a thick initial disc, $Z_\mathrm{d}=0.2$. The bulgeless model has twice the \textcolor{black}{bar} mass in the upper half-space than in the lower at $T \approx 3.5$~Gyr. 
At the same time, the model with a bulge has a completely symmetrical bar evolution! The bulge impact is the same for all models presented: \textcolor{black}{secondary} buckling is damped regardless of the initial disc parameters. \par
\textcolor{black}{As for the first buckling occurrence, Fig.~\ref{fig:asym_big} can be somewhat misleading and here we want to clarify this issue. Not all bulgeless models demonstrate the initial buckling of a bar (see the model with $M_\mathrm{h}=1.5$ and $Z_\mathrm{d}=0.1$). On the other hand, not all their counterparts with bulges \textit{always demonstrate the absence of the first buckling.} For example, the bar in the model with $M_\mathrm{h}=1$ and $M_\mathrm{b}=0.2$ has approximately 20\% asymmetry of the bar mass at $t=1.5$~Gyr. The model with $M_\mathrm{h}=1.5$ and a bulge has even slightly greater amplitude of the bar asymmetry at $1$~Gyr than its bulgeless counterpart (30\% vs 20\%). Although for all other models the presence of the bulge damps the first buckling too. A complete explanation of such inconsistency require a thorough study of initial stages of the bar buckling and \textit{the bending of the stellar disc as a whole} in different models and can lead us too far from the initial subject of the present work. Therefore we omit it for the present and concentrate on study of the prolonged bar buckling which is more important for observational statistics.}
%From this figure follows the main result of the present work.  A typical bulge completely prevents the secondary buckling of a bar. 
%%%%%%%%%%%%%%%%%%%%%%%%%%%%%%%%%%%%%%%%%%%%%%%%%%%%%%%%%%%%%%%%%%%%%%%%%%%%%%%%%%%%%%%%%%%%%%%%%%%%
\section{Boundary mass}
%%%%%%%%%%%%%%%%%%%%%%%%%%%%%%%%%%%%%%%%%%%%%%%%%%%%%%%%%%%%%%%%%%%%%%%%%%%%%%%%%%%%%%%%%%%%%%%%%%%%
\begin{figure}
\begin{minipage}[t]{0.9\textwidth}%
\includegraphics[scale=0.3]{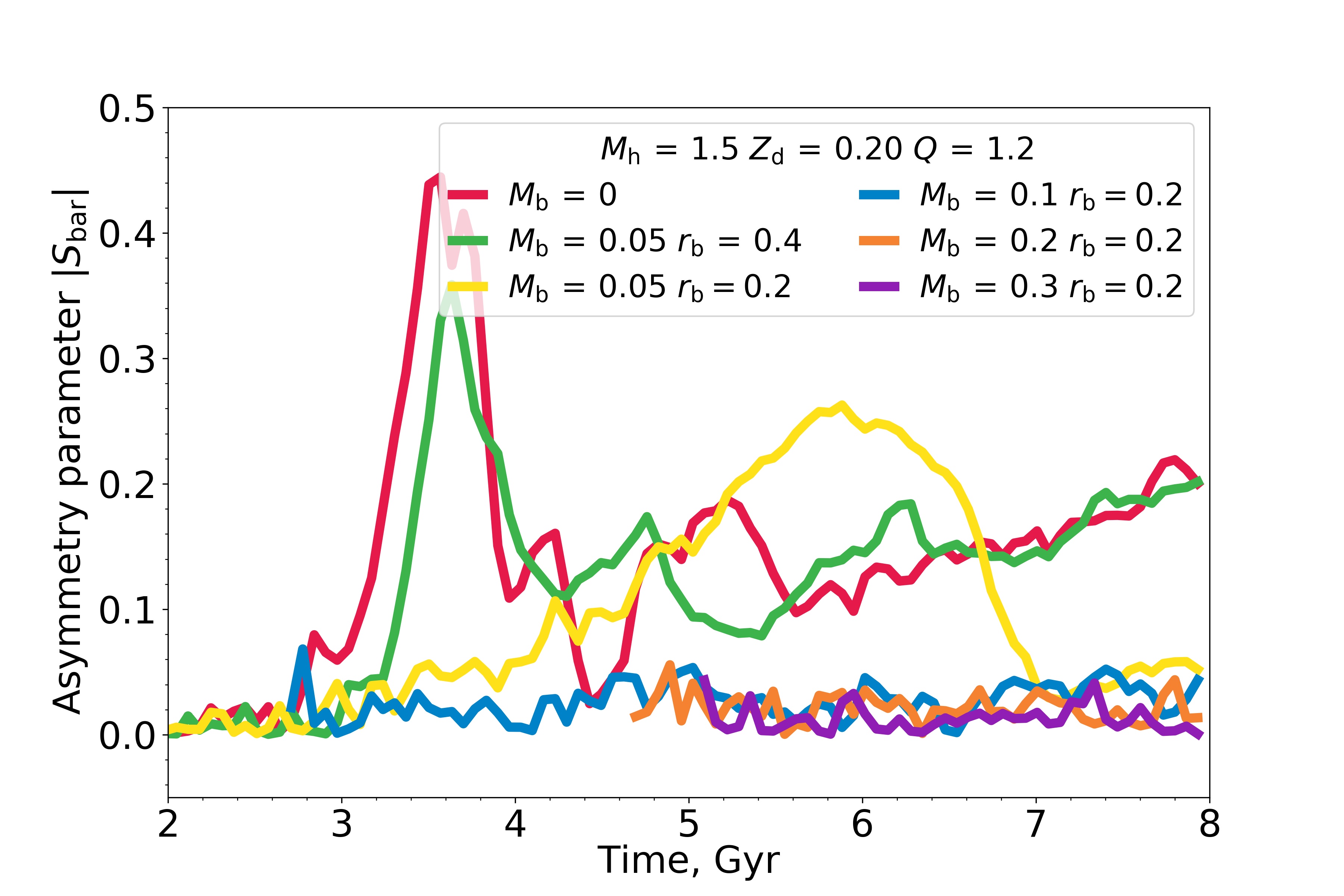}
%\label{fig:axis}
\end{minipage}%
\caption{The same as in Fig~\ref{fig:asym_big} but for models with an extended list of bulge parameters. The limits of the axes are also changed for better readability.}
\label{fig:asym_mass_bound}
\end{figure}
%%%%%%%%%%%%%%%%%%%%%%%%%%%%%%%%%%%%%%%%%%%%%%%%%%%%%%%%%%%%%%%%%%%%%%%%%%%%%%%%%%%%%%%%%%%%%%%%%%%%%%%
The results of the previous section show that a low-mass bulge prevents the secondary buckling of a bar. Therefore, an important question arises: how small should the central concentration be, so that the secondary buckling can be observed. To estimate the boundary, we constructed an additional series of models with bulges of diminishing masses (see Tab.~\ref{tab:models_pars}). As a progenitor (a bulgeless galaxy) of this family of models, we took a model with an initially thick disc $Z_\mathrm{d}=0.2$, because this model demonstrated one of the largest amplitude of the secondary buckling among all other models. Snapshots of the models for different times are shown in Fig.~\ref{fig:model_snapshots}. Fig.~\ref{fig:asym_mass_bound} displays the evolution of the asymmetry parameter for them. For bulge masses $M_\mathrm{b} \geq 0.1$, the \textcolor{black}{secondary} buckling is completely damped ($|S_\mathrm{bar}(t)| \lesssim 0.1$). The model with $M_\mathrm{b}=0.05$ and $r_\mathrm{b}=0.4$ demonstrates the prolonged buckling and the bar is vertically asymmetrical for all times, as in the model without the bulge. The low-mass model with higher central concentration of the bulge ($r_\mathrm{b}=0.2$) show a different behaviour. Although the \textcolor{black}{secondary} buckling occurs at the late stage of the bar evolution, it lasts 2-3 Gyr, and disappears completely by 8 Gyr. Thus, we can conclude that it is not so much the mass that is important as the concentration of the matter, which depends both on $M_\mathrm{b}$ and $r_\mathrm{b}$. In general the model with $M_\mathrm{b}=0.05$ demonstrates that the \textcolor{black}{secondary} buckling may be observed at the present time if the central concentration is small enough, $M_\mathrm{b}<0.1M_\mathrm{d}$ and $r_\mathrm{b}>0.2$. Given that a bulgeless model with $Z_\mathrm{d}=0.2$ shows one of the largest amplitudes of secondary buckling among all models, this result shows that, in general, the mass of a bulge which prevents the secondary buckling of a bar is rather small, smaller than the typical observed values of the bulge mass (e.g., \citealp{Mosenkov_etal2010}). Thus, a typical classical bulge prevents the \textcolor{black}{secondary} buckling.
%(If we consider greater time intervals will there be the secondary buckling or not? Does bulge truly prevent the secondary buckling or just postpone it?)
%%%%%%%%%%%%%%%%%%%%%%%%%%%%%%%%%%%%%%%%%%%%%%%%%%%%%%%%%%%%%%%%%%%%%%%%%%%%%%%%%%%%%%%%%%%%%%%%%%%%
\section{Discussion and Conclusions}
%%%%%%%%%%%%%%%%%%%%%%%%%%%%%%%%%%%%%%%%%%%%%%%%%%%%%%%%%%%%%%%%%%%%%%%%%%%%%%%%%%%%%%%%%%%%%%%%%%%%
We considered a set of models with a wide range of initial stellar disc parameters. In each model the addition of a classical bulge with mild parameters (small mass and a typical scale length) prevents the \textcolor{black}{secondary} buckling of a bar almost independently of initial conditions of the stellar disc.
The result complements the well-known effect of weakening of the bending instability by an excess of central concentration associated with the cooling gas steadily sinking into the centre of the stellar disc \citep{Berentzen_etal1998,Debattista_etal2005}. 
If the latter effect explains why the \textcolor{black}{buckled discs} probably can not be observed in late-type galaxies (where gas makes up a substantial part of the mass), our result explains why the galaxies of early types (which are typically associated with classical bulges) do not demonstrate a vertically asymmetric bar. Thus, the \textcolor{black}{buckling stage} is successfully damped by increasing the central concentration from different sources for both early and late type galaxies.
\par 
Do our results mean that the secondary buckling can not be detected or realised in nature? Our cautious answer is `no' more than `yes'. First, three galaxies of intermediate morphological types with trapezoidal inner isophotes (corresponding to the main buckling region) are suspected of having the ongoing buckling \citep{Erwin_Debattista2017,Li_etal2017}. It is worth noting, however, that one of the galaxies NGC~3227 is interacting with the dwarf elliptical galaxy NGC~3226. Moreover, both galaxies are embedded in a common cocoon of stars and warm gas, which is supposed to be the result of merging of the third galaxy \citep{Appleton_etal2014}. Thus, the shape of the isophotes of NGC~3227 can be greatly distorted due to this interaction. And it is not surprising that one of the two outer-bar spurs of NGC~3227 at the long side of the trapezoid is pointed out exactly to the satellite NGC~3226, which may create a false impression of the position of the spurs. 
Secondly, we have collected some observational evidence that may indirectly point out that the bars in some of the observed galaxies could have passed through the secondary buckling stage (and, therefore, it can be observed too). 
\par
1. Recent studies of S0 galaxies show that an S0 galaxy may have a discy bulge rather than a classic one,
\citep{Vaghmare_etal2013,Vaghmare_etal2018} even a composite structure consisting of a very small classical bulge and a discy pseudobulge \citep{Erwin_etal2015}. 
In some of these galaxies, the HI line is not observed, and the absence of gas opens up the golden road for the \textcolor{black}{secondary} buckling.
Additional observational studies of such galaxies should show whether there is any evidence of bar asymmetry of trapezoidal shape. 
\par
2. \citet{Laurikainen_Salo2017} have recently shown that with a concentrated bulge component, an in-plane bar obtains a barlens morphology. Such a barlens is small, typically 0.5 or less than a bar major axis. In our experiments models with concentrated bulges do not exhibit a prominent primary buckling and do not demonstrate secondary buckling (except one model with a light but concentrated bulge!). At the same time \citet{Laurikainen_Salo2017} have identified a class of galaxies where the barlens contribution dominates in the inner region of the disc and the bar appears only through small spurs near the lens edge (class ``e'' in that work). From perspective of our simulations, such a morphology may be associated with a hot or fairly thick stellar disc (without a classical bulge) where the bar, in turn, is subject to the violent buckling instability \textcolor{black}{at the late stages of the bar evolution} (see~Fig.~12,~13 in \citealp{Smirnov_Sotnikova2018}).
Again, the detailed studies of such galaxies are required to justify the presence of vertical bar asymmetries. 
\par
3. Finally, recent studies of S2B galaxies \citep{Mendez_Abreu_etal2018} have shown that the inner bar of some galaxies also demonstrate the unique X-shaped structure (along with the outer bar and its X-structure), which can be interpreted as a direct consequence of two buckling phases, where the inner X-structure was formed during the first buckling phase and the outer X-structure formed during the \textcolor{black}{secondary} buckling.
\par 
Although in the present work we obtained that a spherical subsystem (a classical bugle) of rather small mass \textcolor{black}{keeps the bar vertically symmetric}, apparently, there may exist conditions under which the total mass of a bulge and gas in the centre is not enough to prevent the \textcolor{black}{secondary} buckling. 
Based on the facts described above, we conclude that it is too early to exclude the possibility of the \textcolor{black}{secondary} buckling and its consequences for the dynamics of real galaxies. \textcolor{black}{But if ongoing buckling is not found among objects listed above, it will create a  controversial situation between numerical simulations and observations, challenging to look for other reasons besides gas and bulges, that force bars to be  unbuckled.}

%%%%%%%%%%%%%%%%%%%%%%%%%%%%%%%%%%%%%%%%%%%%%%%%%%%%%%%%%%%%%%%%%%%%%%%%%%%%%%%%%%%%%%%%%%%%%%%%%%%%
\section*{Acknowledgements}
%%%%%%%%%%%%%%%%%%%%%%%%%%%%%%%%%%%%%%%%%%%%%%%%%%%%%%%%%%%%%%%%%%%%%%%%%%%%%%%%%%%%%%%%%%%%%%%%%%%%%%%
The authors express gratitude for the grant of the Russian Foundation for Basic Researches number 19-02-00249.
We also thank the anonymous referee for the review and appreciate the comments. We also highly appreciate the constructive comments and suggestions made by the scientific editor, which allowed us to improve the quality of the publication.

\bibliographystyle{mnras}
\bibliography{bulge_vs_bend}
%%%%%%%%%%%%%%%%%%%%%%%%%%%%%%%%%%%%%%%%%%%%%%%%%%%%%%%%%%%%%%%%%%%%%%%%%%%%%%%%%%%%%%%%%%%%%%%%%%%%%%%

%%%%%%%%%%%%%%%%%%%%%%%%%%%%%%%%%%%%%%%%%%%%%%%%%%%%%%%%%%%%%%%%%%%%%%%%%%%%%%%%%%%%%%%%%%%%%%%%%%%%
\end{document}